\documentclass{SCIS2023}
\usepackage{multirow}
\usepackage[T1]{fontenc}
\usepackage{epstopdf}
\usepackage{makecell}
\usepackage{soul}
\usepackage{url}
\usepackage{enumitem}

\begin{document}
\ArticleType{RESEARCH PAPER}
\Year{2024}
\Month{}
\Vol{}
\No{}
\DOI{}
\ArtNo{}
\ReceiveDate{}
\ReviseDate{}
\AcceptDate{}
\OnlineDate{}

\newcommand{\shuqi}[1]{\textcolor{magenta}{#1}}
\newcommand{\rem}[1]{\textcolor{red}{\st{#1}}}

\title{
A universal optimization framework based on cycle ranking for influence maximization in complex networks
%A general framework for finding multiple influential nodes in complex networks based on cycle ranking
}

\author[1]{Wenfeng Shi}{}
\author[1]{Tianlong Fan}{tianlong.fan@ustc.edu.cn}
\author[2]{Shuqi Xu}{xushuqi@idata.ah.cn}
\author[3]{Rongmei Yang}{}
\author[1,2,3]{Linyuan L\"u}{linyuan.lv@ustc.edu.cn}

\AuthorMark{Shi W F}

\AuthorCitation{Shi W F, Fan T L, Xu S Q, et al}

\address[1]{School of Cyber Science and Technology, University of Science and Technology of China, \\Hefei {\rm 230026}, China}
\address[2]{Institute of Dataspace, Hefei Comprehensive National Science Center, Hefei {\rm 230088}, China}
\address[3]{Institute of Fundamental and Frontier Sciences, University of Electronic Science and Technology of China, \\Chengdu {\rm 611731}, China}

\abstract{Influence maximization aims to identify a set of influential individuals, referred to as influencers, as information sources to maximize the spread of information within networks, constituting a vital combinatorial optimization problem with extensive practical applications and sustained interdisciplinary interest. Diverse approaches have been devised to efficiently address this issue, one of which involves selecting the influencers from a given centrality ranking. In this paper, we propose a novel optimization framework based on ranking basic cycles in networks, capable of selecting the influencers from diverse centrality measures. The experimental results demonstrate that, compared to directly selecting the top-$k$ nodes from centrality sequences and other state-of-the-art optimization approaches, the new framework can expand the dissemination range by 1.5 to 3 times. Counterintuitively, it exhibits minimal hub property, with the average distance between influencers being only one-third of alternative approaches, regardless of the centrality metrics or network types. Our study not only paves the way for novel strategies in influence maximization but also underscores the unique potential of underappreciated cycle structures.
%171words

}


\maketitle

\section{Introduction}
The overall performance of complex networks, both structurally and dynamically, is often determined by a specific set of nodes. These nodes, despite their small number, wield a disproportionately large influence, rapidly disseminating information throughout social networks~\cite{kempe2003maximizing}, disrupting the structure or functionality of infrastructure networks~\cite{ren2019generalized}, and effectively preventing the spread of pandemics~\cite{clusella2016immunization,altarelli2014containing}. Identifying this set of nodes with the greatest overall impact, termed ``influencers’’~\cite{morone2015influence}, has long been one of the most crucial issues in network science, garnering continuous interdisciplinary attention.

The identification of influencers has been proven to be an NP-hard problem~\cite{kempe2003maximizing} that cannot be directly and precisely solved. A profusion of approaches from various perspectives have been proposed, categorized into five categories:

\begin{itemize}[leftmargin=2em]
\item[i)] Monte Carlo simulation-based methods~\cite{kempe2003maximizing, leskovec2007cost, goyal2011data,chen2010scalable,borgs2014maximizing}. These methods typically rely on greedy strategies and computationally intensive simulations, often failing to achieve global optimality and unsuitable for large-scale networks;
\end{itemize}

\begin{itemize}[leftmargin=2em]
\item[ii)] methods based on centrality and network topology. These methods seek influencers based on node centrality measuress~\cite{chen2009efficient,lu2016vital} or network topological features~\cite{shang2017cofim,li2020community}. They involve direct selection of the top-$k$ nodes with the highest centrality values or considering their dispersibility, selecting according to certain rules. Such methods are concise and effective, allowing for the selection of appropriate centrality measures tailored to specific topological features of the target network, such as degree centrality for high heterogeneity and bridgeness centrality~\cite{jensen2016detecting} for community structure, etc., exhibiting wide applicability;
\end{itemize}

\begin{itemize}[leftmargin=2em]
\item[iii)] methods based on optimization theory. These methods map the influence maximization problem into an operable global optimization objective. Optimal percolation~\cite{morone2015influence}, message passing~\cite{altarelli2014containing, braunstein2016network}, belief propagation~\cite{altarelli2013optimizing, mugisha2016identifying}, etc., fall into this category, offering better results than Monte Carlo methods and typically requiring time roughly linearly correlated with the size of the network edges or nodes. However, such methods often rely on numerous approximations and simplifications, hence typically providing only approximate solutions. Furthermore, influence maximization propagation and maximization blocking are often not equivalent~\cite{radicchi2017fundamental}, and the optimization objectives lack intuitive interpretation for specific contexts; 
\end{itemize}

\begin{itemize}[leftmargin=2em]
\item[iv)] methods based on percolation. These methods identify influencers by considering nodes near critical points that can trigger dramatic changes in global connectivity~\cite{ji2017effective,schneider2012inverse,kolumbus2021influence}, especially during explosive percolation processes~\cite{clusella2016immunization,qiu2021identifying,peng2023unveiling}. Such methods are easy to understand, exhibit multiple solutions and high noise tolerance, and because the clusters formed during percolation are universal, they do not depend on specific network attributes such as network size, sparsity, degree distribution, and heterogeneity;
\end{itemize}

\begin{itemize}[leftmargin=2em]
\item[v)] methods based on machine learning~\cite{fan2020finding,kumar2022influence,ling2023deep,li2023survey}. These methods assess node influence through learning from large-scale data, effectively identifying influencers. Research in this field is of significant importance for social network analysis, as it can better handle large-scale, high-dimensional data and unearth deeper patterns and regularities, with excellent scalability.
\end{itemize}

\noindent Additionally, there are further achievements~\cite{li2018influence,banerjee2020survey,singh2022influence}. As the demands and scale of data increase, enhancing the efficiency, accuracy, and interpretability of algorithms, addressing specific issues in various backgrounds and network types, and integrating them with other social network data mining tasks, are emerging as current trends.

Focusing on methods in category (ii), several optimization frameworks have been developed to reduce the overlap in their influence ranges compared to directly selecting the top-$k$ nodes with the highest centrality values~\cite{shi2023cost,lu2011leaders,fan2021characterizing}, denoted as \emph{TopK}. Three representative ones are as follows: 1) Avoiding direct connections among influencers~\cite{he2015novel,kitsak2010identification} (referred to as \emph{NotCon}); 2) Selecting candidates that increase the average distance between them~\cite{shi2023cost,cui2023research,tao2022identifying,hu2014effects} (referred to as \emph{IncDis}); 3) Selecting candidates that decrease the average similarity between them~\cite{masterShi} (referred to as \emph{DecSim}). These optimization frameworks inherently seek a delicate trade-off between the individual influence of influencers and their overall dispersibility, effectively enhancing the influence range of influencers.

Recently, cycle structures have garnered widespread attention due to their connectivity redundancy in structure and feedback effects in dynamics, and they are abundant in networks~\cite{bianconi2003number}. They play a crucial role in the formation of local clusters~\cite{lizier2012information}, communities~\cite{radicchi2004defining}, robustness~\cite{braunstein2016network} and stability~\cite{neutel2002stability} of networks. Without cycles, networks degenerate into trees, resulting in the loss of any emergent structures and behavioral complexity. Different cycles have been shown to play a vital role in processes such as the global distribution and integration of information in the brain~\cite{sizemore2018cliques}, as well as network decomposition processes~\cite{zhou2013spin,morone2015influence, mugisha2016identifying, peng2023unveiling}. Several centrality metrics based on cycle structures have been proposed. The cycle ratio identifies nodes that exhibit high influence across multiple dynamics by considering the extent to which nodes participate in the smallest cycle among their cycle neighbors~\cite{fan2021characterizing}. Another metric based on basic cycles, the cycle number, also demonstrates excellent performance due to its inherent global dispersibility and economy~\cite{fan2019towards,shi2023cost}. On the other hand, methods have been proposed to measure the importance of cycles themselves, revealing that critical cycles significantly impact the synchronization capabilities of networks~\cite{jiang2023searching}. Further deepening our understanding of the important role played by cycles and strengthening their utilization is imperative.

In this paper, we propose a novel optimization framework, referred to as \emph{CycRak}, for selecting influencers from the top-$k$ nodes based on centrality, achieved through the ranking of basic cycles. Our underlying idea here is that influencers are located on the most important basic cycles, where the most important basic cycles are those that span more communities at the macroscopic level, involve more efficient paths at the mesoscopic level, and are more likely to articulate different local dense centers at the microscopic level. Specifically, we first rank the importance of basic cycles and then select, one by one, a node with the highest centrality from each cycle as an influencer, under the premise that the selected influencers are not connected to each other. Here, six different node centrality measures are considered.

The experimental results on empirical and synthetic networks demonstrate that, compared to four benchmark frameworks, namely TopK, NotCon, IncDis, and DecSim, the influencers identified by the CycRak framework exhibit the best propagation performance, with a dissemination range 1.5 to 3 times that of the benchmark frameworks. Furthermore, these influencers demonstrate more significant global dispersibility, i.e., larger average distances between them, and counterintuitively, they possess the lowest hub characteristics, with their average degree being only one-third of the benchmark frameworks. These results illustrate that the CycRak framework not only achieves a delicate balance between node influence and dispersibility but also enhances penetrability across dimensions of influence while maintaining the best economy.

\section{Methods}

\subsection{Cycle-ranking based framework for identifying influencers (CycRak)}

Consider an undirected, unweighted network $G(V, E)$ that disallows self-loops and parallel edges, where $V$ represents the set of nodes with $|V| = N$, and $E$ represents the set of edges with $|E| = M$. Due to the incalculable number of all cycles and the minimal impact exerted by a vast majority of longer cycles on dynamics~\cite{fan2021characterizing}, we focus solely on a specific type of cycle here, namely, basic cycles, also known as fundamental cycles. Basic cycles denoted as $B = \{b_1,b_2,\ldots,b_{M-N+1}\}$, constitute a set of simple cycles forming a basis for the cycle space of the network, with their quantity being $M-N+1$ (when the network is connected). Each of them is linearly independent from the remaining cycles, and any other cycle in the network can be constructed from basic cycles through symmetric difference operations. Therefore, they provide a holistic view of the network's cycle structure~\cite{shi2023cost}. Basic cycles can be readily captured from any spanning tree of the given network, hence exhibiting multiple solutions.

Given the variable sizes of cycles, which can constitute small local structures or span multiple local regions of a network due to their larger sizes, their primary influence may stem from the cycles as a whole or from the contributions of the nodes or edges comprising them. Consequently, the importance of a cycle is determined by multiple dimensions. Here, we quantify the importance of basic cycles from three different perspectives. Specifically, at the macroscopic level, we design an indicator, $I_{com}$, to measure the extent to which a basic cycle participates in communities,
\begin{equation}
I_{com} = \frac{N_{c}}{n} \times \frac{N_{cn}}{n_{cn}},\label{eq1}
\end{equation}
where $N_{c}$ denotes the sum of the number of communities involved by $n$ nodes in a basic cycle $b$, $N_{cn}$ represents the sum of the number of communities involved by neighbors of $b$'s nodes, and $n_{cn}$ denotes the number of these neighbors. A larger $I_{com}$ indicates that a cycle is associated with more communities. Here, we employ the efficient and widely used Louvain algorithm~\cite{blondel2008fast} to achieve community detection in the network.

At the mesoscopic level, we introduce an indicator $I_{pth}$ to measure the number of effective paths in which a basic cycle participates,
\begin{equation}
I_{pth} = \frac{\sum_{e \in b}{\eta_e}}{m},\label{eq2}
\end{equation}
where $\eta_e$ denotes the current-flow betweenness centrality~\cite{brandes2005centrality} of edge $e$ in $b$, and $m$ represents the number of edges in $b$ (which is actually equal to $n$). In detail,

\begin{equation}
\eta_e = \frac{1}{(N-1)(N-2)}\sum_{s \neq t\in V}{\tau_{st}{(e)}},\label{eq2.2}
\end{equation}
where $\tau_{st}{(e)}$ denotes the throughput in the case of an st-current, which is the current flowing from a source node $s$ to a target node $t$. Unlike traditional betweenness centrality~\cite{brandes2001faster}, which considers only the shortest paths, current-flow betweenness centrality also takes into account other shorter paths. This measure, also commonly referred to as random-walk betweenness centrality~\cite{newman2005measure}, emphasizes the role of edges in the actual flow of electrical current, providing a better reflection of the edges' real influence within the network. A larger $I_{pth}$ for a cycle indicates its involvement in more effective paths, thus rendering it more crucial at the mesoscopic level. 

At the microscopic level, we introduce an indicator $I_{lc}$ to measure the ability of a basic cycle to articulate different local dense topological centers,
\begin{equation}
I_{lc} = \frac{\sum_{i \in b}{N_i}}{n},\label{eq3}
\end{equation}
where $N_i$ represents the number of basic cycles involving node $i$ in basic cycle $b$, i.e., the number of basic cycles in $B$ that contain node $i$. Local dense centers in the network, such as local clusters, consist of multiple trivial cycles, often requiring the involvement of several basic cycles for their formation. Therefore, when nodes within a basic cycle participate in more other basic cycles, it reflects that the basic cycle articulates more diverse local dense centers.

\begin{figure}[htbp]
\centering
\includegraphics[width=\textwidth, keepaspectratio]{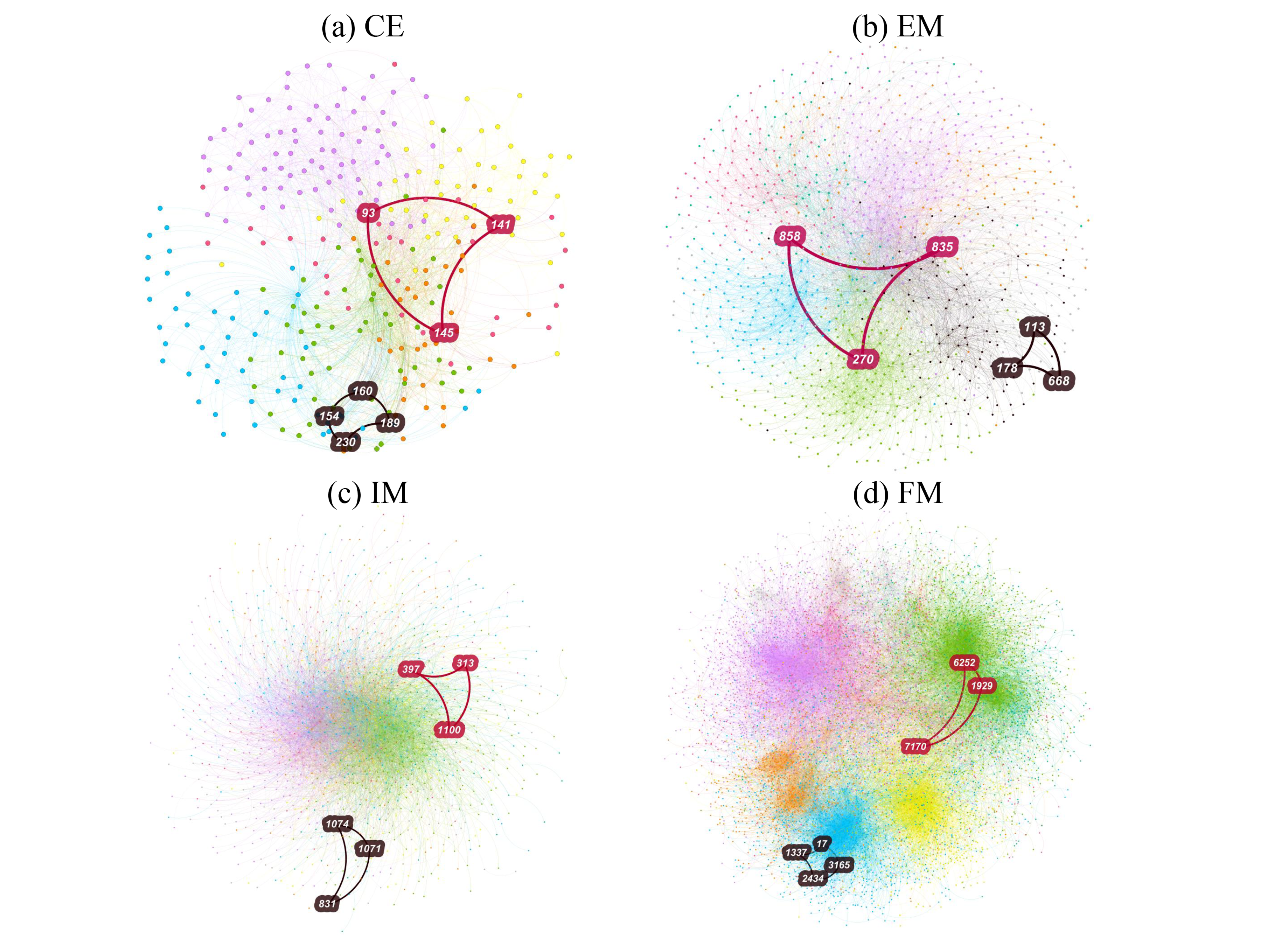}
\caption{Visualization of the most important (in red) and least important (in dark brown) basic cycles in four empirical networks. Different communities within the networks are identified using the Louvain algorithm, distinguished by the colors of nodes and edges.}
\label{fig1}
\end{figure}

Finally, by integrating the contributions of importance from the three levels, characterized respectively from the perspectives of communities, edges, and nodes, the overall importance $I_b$ of a basic cycle $b$ is defined as:
\begin{equation}
I_b = I_{com}*I_{pth}*I_{lc}.\label{eq4}
\end{equation}

Figure ~\ref{fig:s1} in \nameref{sec:SI} illustrates the relationship between the importance of basic cycles, as captured by Eq.~\ref{eq4}, and their lengths across four empirical networks (see Section ~\ref{Sec:data} for details). This relationship is a significant yet unanswered question, as previous studies have either focused solely on specific lengths of cycles, such as triangles~\cite{jiang2023searching}, or confined their analysis to the shortest cycles~\cite{fan2021characterizing}. Notably, the relationship between cycle importance and length exhibits a consistent pattern across diverse networks, namely, the most important cycles tend to be the shortest, while longer cycles display diminished importance. This pattern confirms previous research~\cite{fan2021characterizing}. Furthermore, cycles of equal length exhibit significant variations in importance, with the variance in importance being more pronounced among shorter cycles. This underscores the notion that short cycles do not invariably equate to importance, underscoring the imperative of rigorously quantifying their importance.

Figure ~\ref{fig1} intuitively illustrates the most and least important basic cycles in these four empirical networks. It can be observed that all the most important cycles indeed have the shortest lengths and tend to span across different communities, as shown in Figures ~\ref{fig1}a and ~\ref{fig1}b with three distinct communities, and Figures ~\ref{fig1}c and ~\ref{fig1}d with two communities. Furthermore, they engage in more efficient pathways, demonstrating strong control over the flow of information in the global network. These cycles also involve nodes that articulate the most local dense clusters, as clearly demonstrated in the results presented in Table ~\ref{tab:s1} in \nameref{sec:SI}. In contrast, the least important cycles tend to reside within communities, potentially longer, and participate in the construction of fewer efficient pathways and local articulations.

The proposed CycRak framework comprises two stages. In the first stage, basic cycles are sorted based on their importance $I_b$. The second stage involves selecting influencers from the most important basic cycle onwards, one at a time, by choosing the node with the highest centrality value in each cycle, under the constraint that the candidate is not directly connected to any existing influencer. This selection process terminates when a total of $k$ influencers have been obtained. Specifically, at each step, the framework selects the node with the highest centrality from a basic cycle, excluding nodes already chosen as influencers. It then checks if the selected node satisfies the connectivity constraint. If so, it is added as a new influencer; otherwise, the process moves to the next basic cycle. In cases where multiple nodes simultaneously satisfy the constraint, one is randomly selected as the influencer. The complete pseudocode for the CycRak framework can be found in Algorithm \ref{alg1} in \nameref{sec:SI}. Here, the centrality of nodes can theoretically be any type, chosen flexibly according to specific requirements. This study considers six centralities, as detailed in Table \ref{tab1}.

\subsection{Benchmark frameworks}
We consider four benchmark frameworks for identifying $k$ influencers in complex networks, all designed based on node centralities, namely \emph{Top$K$}~\cite{shi2023cost,lu2011leaders,fan2021characterizing}, \emph{NotCon}~\cite{he2015novel, kitsak2010identification}, \emph{AvgDis}~\cite{shi2023cost, cui2023research, tao2022identifying,hu2014effects}, and \emph{AvgSim}~\cite{masterShi}, respectively. We denote the node sequence sorted by node centrality as $V_R = (v_1, v_2, \ldots, v_n)$ and use $S$ to represent the set of influencers. $S = \varnothing$, initially. The definitions of these four optimization frameworks are as follows:

\textbf{\emph{TopK}}: Directly selecting the top-$k$ nodes as influencers from $V_R$, denoted as $S = \{v_1, v_2, \ldots, v_k\}$.

\textbf{\emph{NotCon}}: Sequentially selecting non-connected nodes from $V_R$. If a node $v_i$ is connected to any influencer in $S$, skip it. Stop when $|S|=k$.

\textbf{\emph{IncDis}}: The average distance between influencers has been demonstrated to significantly influence the extent of diffusion. Generally, the larger this average distance, the more dispersed the influencers, consequently resulting in a broader diffusion range. Specifically, nodes from $V_R$ are sequentially selected to join $S$ if their inclusion increases the average distance $d$ between influencers. Stop when $|S|=k$. Here, $d$ is defined as 
\begin{equation}
d = \frac{2\sum_{i \in S}{\sum_{j \in S\backslash i}{d_{ij}}}}{|S|(|S| - 1)},
\end{equation}
where $d_{ij}$ represents the shortest path length between node $i$ and $j$.

\textbf{\emph{DecSim}}: The average similarity among influencers also significantly influences the extent of diffusion, as it characterizes the degree of overlap in the diffusion ranges of influencers. A smaller average similarity implies fewer common neighbors among influencers, resulting in a greater number of different targets they have the potential to infect. Specifically, nodes from $V_R$ are sequentially selected to join $S$ if their inclusion decreases the average similarity $s$ between influencers. Stop when $|S|=k$. Here, $s$ is defined as
\begin{equation}
s = \frac{2\sum_{i \in S}{\sum_{j \in S\backslash i}{s_{ij}}}}{|S|(|S| - 1)},
\end{equation}
where $s_{ij}$ represents the Jaccard similarity coefficient~\cite{jaccard1912distribution} between nodes $i$ and $j$, and $s_{ij}$ is
\begin{equation}
s_{ij} = \frac{|\Gamma (i) \cap \Gamma (j)|}{|\Gamma (i) \cup \Gamma (j)|},
\end{equation}
where $\Gamma (i)$ represents the neighbor set of node $i$.

\subsection{Node Centralities}\label{Sec:centr}

We provide six representative node centralities for ranking nodes, including Degree centrality (DC) ~\cite{newman2018networks}, H-index centrality (HC)~\cite{lu2016h}, Semi-local centrality (LC)~\cite{chen2012identifying}, and Collective influence (CI) ~\cite{morone2016collective} that utilize limited local topological information, as well as Eigenvector centrality (EC)~\cite{bonacich1972factoring} and Closeness centrality (CC)~\cite{freeman1978centrality} that utilize global topological information. Table \ref{tab1} provides a brief description and definition of these centralities.

\begin{table}[H]
\footnotesize
\caption{Description and definition of the six centralities.}
\label{tab1}
\tabcolsep 6pt
\renewcommand\arraystretch{2.5}
\begin{center}
\begin{tabular*}{\textwidth}{p{3.7cm}p{6.8cm}p{4cm}}%{p{2.5cm}p{9cm}p{}}%{@{}lll@{}}
\toprule           

    Centrality & \makecell[c]{Description}     & \makecell[c]{Definition}  \\
    \midrule
    
    \textbf{\emph{Degree (DC)}} & Importance determined by the number of neighbors.   & $DC_i = \sum_{j = 1}^N{A_{ij}}$   \\

    \textbf{\emph{H-index (HC)}} & Importance determined by the number of neighbors and their importance.  & $HC_i = \mathcal{H}(k_{j_1}, k_{j_2}, \ldots, k_{j_k})$  \\

    \textbf{\emph{Semi-local centrality (LC)}} & Importance determined by information from neighbors within four-orders.  & $LC_i = \sum_{j \in \Gamma_i}{\sum_{u \in \Gamma_j}{N(u)}}$  \\

    \textbf{\emph{Collective influence (CI)}} & Importance determined by the number of neighbors and the degrees of $l$-th order neighbors.  & $CI_i = (k_i - 1)\sum_{d_{ij}=l}{(k_j - 1)}$  \\

    \textbf{\emph{Eigenvector centrality (EC)}} & Importance determined by the number of neighbors and their importance, with the importance of neighbors captured iteratively using global topological information. & $EC_i = q_i = c\sum_{j = 1}^N{A_{ij}q_j}$\\
    
    \textbf{\emph{Closeness (CC)}} & Importance determined by the average distance from the node to all other nodes, i.e., closer to the geometric center of the network is more important.  & 
    $CC_i = \frac{1}{N - 1}\sum_{j (\neq i)}\frac{1}{d_{ij}}$ \\
    
    \bottomrule
\end{tabular*}
\end{center}
\end{table}

\subsection{Propagation model}

To quantify the diffusion capability of influencers, we simulated Susceptible-Infected-Recovered (SIR) propagation dynamics~\cite{pastor2015epidemic}, where each node is in one of three states: susceptible, infected, or recovered. At each time step, each infected node infects each of its susceptible neighbors with probability $\gamma$, and then transitions to the recovered state with probability $\mu$. Initially, the influencers determined $S$ by a given scheme are set as infected, while the rest are susceptible. The dynamics terminate when there are no more infected nodes, at which time we calculate the ratio of recovered nodes $N_R$ to the total number of nodes $N$, denoted as $F = N_R/N$, to represent the influence of the influencers. We consider multiple values of $\gamma$ where $\gamma > \beta_c$, with $\mu = 1$. Here, $\beta_c$ is the propagation threshold, given by $\beta_c =$ $\langle k \rangle$/($\langle k^2 \rangle$ - $\langle k \rangle$)~\cite{pastor2015epidemic}, where $\langle k \rangle$ and $\langle k^2 \rangle$ are the average degree and the average squared degree, respectively.

\section{Datasets}\label{Sec:data}

To evaluate the performance of the proposed CycRak framework in maximizing influence, we conducted SIR propagation simulation experiments on four empirical networks from disparate fields and multiple synthetic networks. These four empirical networks are: the C. elegans network~\cite{watts1998collective}, a neural network of the Caenorhabditis elegans nematode, where nodes represent neurons and edges represent synapses; the Email network~\cite{guimera2003self}, depicting email exchanges among members of the Rovira i Virgili University in Spain, with nodes representing users and bidirectional communication represented by edges; the Ia-fb-messages network~\cite{networkrepository}, a Facebook-like social network derived from an online community for students at the University of California, Irvine, comprising users who have sent or received at least one message; and the Asian-last.fm network~\cite{rozemberczki2020characteristic}, a social network of Last.FM users in Asia, illustrating mutual following relationships among users. Our synthetic networks encompass three representative models: Barabási–Albert (BA) networks ~\cite{barabasi1999emergence}, Watts–Strogatz (WS) networks~\cite{watts1998collective}, and Erdős–Rényi (ER) networks~\cite {erdHos1960evolution}. For BA networks, in the process of network generation, each iteration, a new node is introduced and connected to 5 (BA1), 4 (BA2), and 3 (BA3) existing nodes, respectively, following the preferential attachment rule~\cite{newman2001clustering}. For WS networks, we first initialize a nearest-neighbor coupling network, in which each node is connected with its six nearest neighbors, the rewiring probabilities for each edge are 0.08 (WS1), 0.05 (WS2), and 0.02 (WS3), respectively. For ER networks, we set the probability of connecting any two nodes as 0.03 (ER1), 0.02 (ER2), and 0.01 (ER1), respectively. Table~\ref{tab2} provides a detailed overview of the basic topological statistics and the propagation thresholds of the SIR model for these 13 networks.

\begin{table}[H]
\footnotesize
\caption{Topological statistics and the propagation threshold of the SIR model for the four empirical networks and nine synthetic networks. These statistics include the number of nodes ($N$), the number of links ($M$), the average clustering coefficient ($C$), density ($D$), average degree ($\langle k \rangle$), and modularity ($Q$). $\beta_c$ denotes the propagation threshold of the SIR model.}
\label{tab2}
\tabcolsep 12pt
\begin{center}
\begin{tabular*}{\textwidth}{@{}lcccccccc@{}}
\toprule
    Networks & Abbreviation  & $N$     & $M$     & $C$     & $D$     & $\langle k \rangle$    & $Q$ & $\beta_c$\\
    \midrule
    C. elegans & CE & 297   & 2148  & 0.292  & 0.049 & 14.464   & 0.403 & 0.040 \\
    Email & EM &1133  & 5451 & 0.220  &0.009 & 9.622    & 0.571 & 0.057\\
    Ia-fb-messages & IM &1266  & 6451 & 0.068  & 0.008 & 10.191   & 0.310 & 0.038  \\
    Asian-last.fm & FM &7624  & 27806 & 0.219  & 0.001 & 7.294   & 0.816  & 0.041\\

    BA\_network1 & BA1 & 3000  & 14975 & 0.017  &0.003 & 9.983    & 0.281 & 0.044\\
    BA\_network2 & BA2 & 3000  & 11984 & 0.015  & 0.003 & 7.989   & 0.319  & 0.055\\     
    BA\_network3 & BA3 & 3000   & 8991  & 0.012  & 0.002 & 5.994   & 0.388  & 0.068\\
    
    % 重连概率0.02
    WS\_network1 & WS1 & 3000   & 9000  & 0.560  & 0.002 & 6.000   & 0.929  & 0.199\\
    % 重连概率0.05
    WS\_network2 & WS2 & 3000  & 9000 & 0.519  & 0.002 & 6.000   & 0.909   & 0.198\\
    % 重连概率0.08
    WS\_network3 & WS3 & 3000  & 9000 & 0.472  &0.002 & 6.000    & 0.879 & 0.197\\

    % 连边概率0.03
    ER\_network1 & ER1 & 1000  & 15009 & 0.030  &0.030 & 30.018    & 0.161 & 0.033\\
    % 连边概率0.02
    ER\_network2 & ER2 & 1000  & 10002 & 0.020  &  0.020 & 20.004   & 0.201   & 0.050\\
    % 连边概率0.01
    ER\_network3 & ER3 & 1000   & 5070  & 0.009  & 0.010 & 10.140   & 0.275 & 0.099\\

    \bottomrule
\end{tabular*}
\end{center}
\end{table}

\section{Results}

We conducted comprehensive tests to analyze the performance of the CycRak framework in various scenarios and parameter spaces, comparing it with benchmark frameworks to demonstrate its superiority. In simulations, we first captured the descending node rankings $V_R$ for each of the six centralities. Subsequently, we applied CycRak and four benchmark optimization frameworks to select the influencers $S$. Consequently, for each network, a total of $6 \times 5 = 30$ influencer selection schemes were obtained. Given the extensive volume of results and the convenience of discussion, we present the results separately for empirical and synthetic networks. In addition, to provide a clearer depiction of the percentage of top-$k$ influencers, we use top-$\rho$ throughout the subsequent discussions, where $\rho=k / N \times 100\%$.

\subsection{Empirical networks}

\begin{figure}[htbp]
\centering
\includegraphics[width=\textwidth, keepaspectratio]{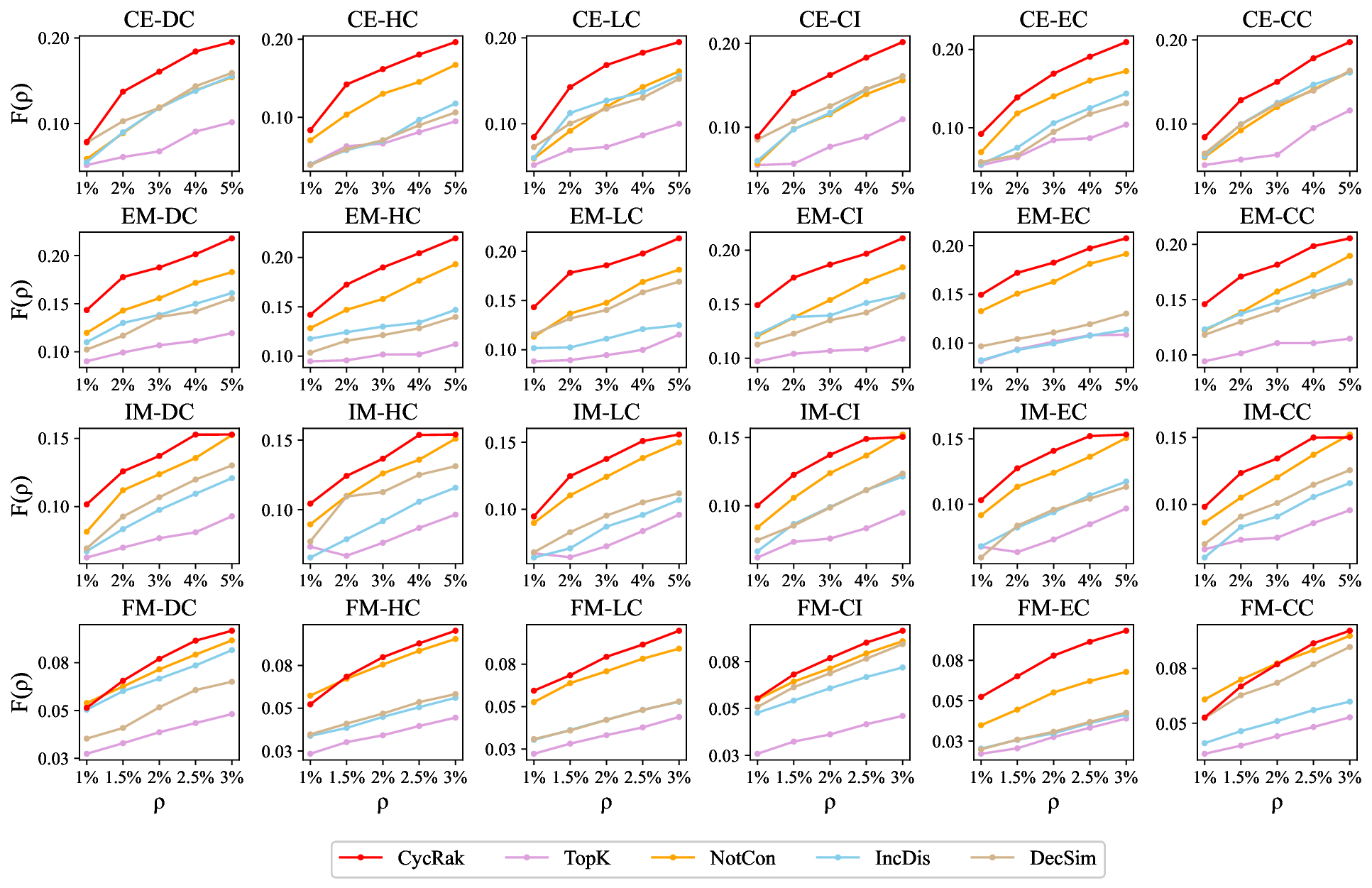}
\caption{Comparison of the influence with different percentages of influencers on four empirical networks. Each panel presents results for a specific centrality, with the network and centrality name indicated in its title. Each curve depicts the collective influence $F(\rho)$ of influencers selected by a particular optimization framework as a function of their percentage $\rho$. These results are captured under SIR model parameters $\gamma=1.25\beta_c$ and $\mu=1$, and each ($\rho$,$F(\rho)$) represents the average of 300 independent realizations.}
\label{fig2}
\end{figure}

Figure~\ref{fig2} illustrates the performance of CycRak and benchmark optimization frameworks across four empirical networks. It can be observed that in all $6 \times 4 = 24$ sets of results, CycRak consistently outperforms other methods, demonstrating significantly superior performance in most cases than the suboptimal solutions, and displaying stable maximal influence, largely unaffected by $\rho$, centrality, or network characteristics. Conversely, the influence achieved by TopK is consistently minimal, validating the effectiveness and necessity of alternative optimization frameworks. Moreover, NotCon exhibits suboptimal performance in most scenarios, highlighting its competitiveness and the potential for substantial gains from its straightforward concept.

\begin{figure}[htbp]
\centering
\includegraphics[width=0.7\textwidth, keepaspectratio]{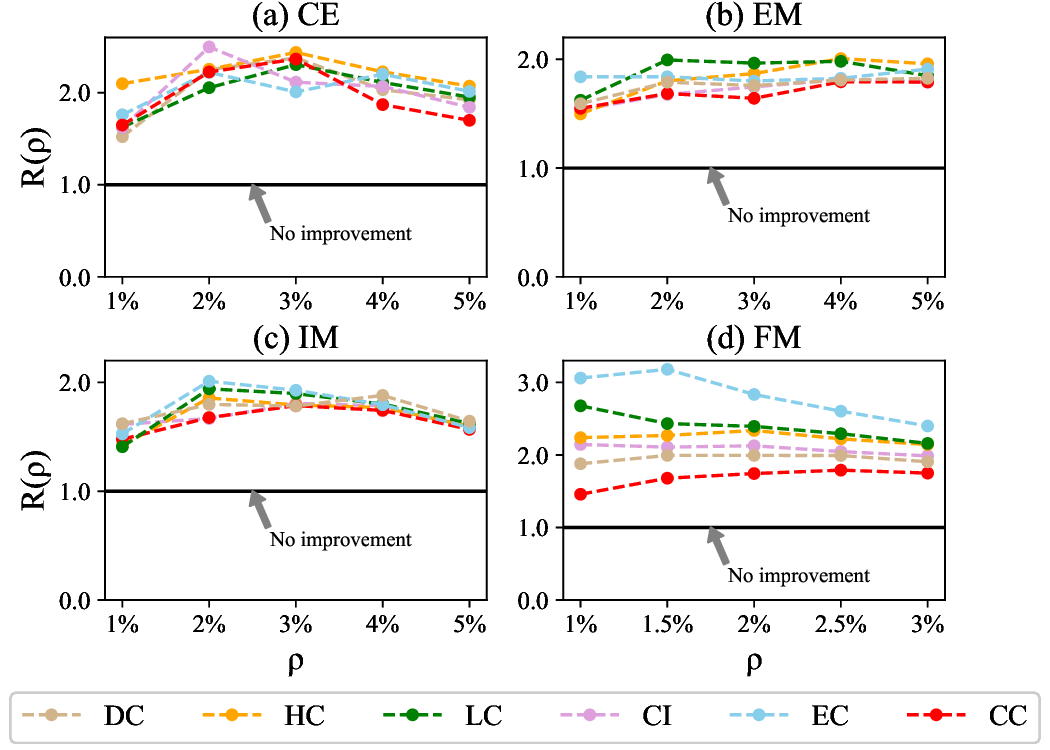}
\caption{Performance improvement of CycRak relative to TopK with different percentages of influencers for four empirical networks. Here $\rho$ represents the percentage of influencers, and $R(\rho)$ indicates the factor by which CycRak's performance exceeds that of TopK. The black solid line denotes the influence of TopK, serving as the baseline, with each dashed line corresponding to a centrality.}
\label{fig3}
\end{figure}

Figure~\ref{fig3} further quantifies the performance improvement brought by CycRak over TopK as depicted in Figure~\ref{fig2}. We define a ratio $R(\rho)$ to represent the influence of CycRak relative to TopK, given by $R(\rho) = F(\rho)_{CycRak} / F(\rho)_{TopK}$. A larger $R(\rho)$ indicates a more significant performance enhancement by CycRak, visually represented by the colored dashed lines diverging further from the black baseline representing TopK ($R(\rho)=1$). Across the four empirical networks, CycRak consistently enhances the diffusion capability of TopK to a level 1.5 to 3 times higher than the latter across various influencer percentages for every centrality metric, with the most notable effect observed in the EC metric of the FM network.

\begin{figure}[!ht]
\centering
\includegraphics[width=\textwidth, keepaspectratio]{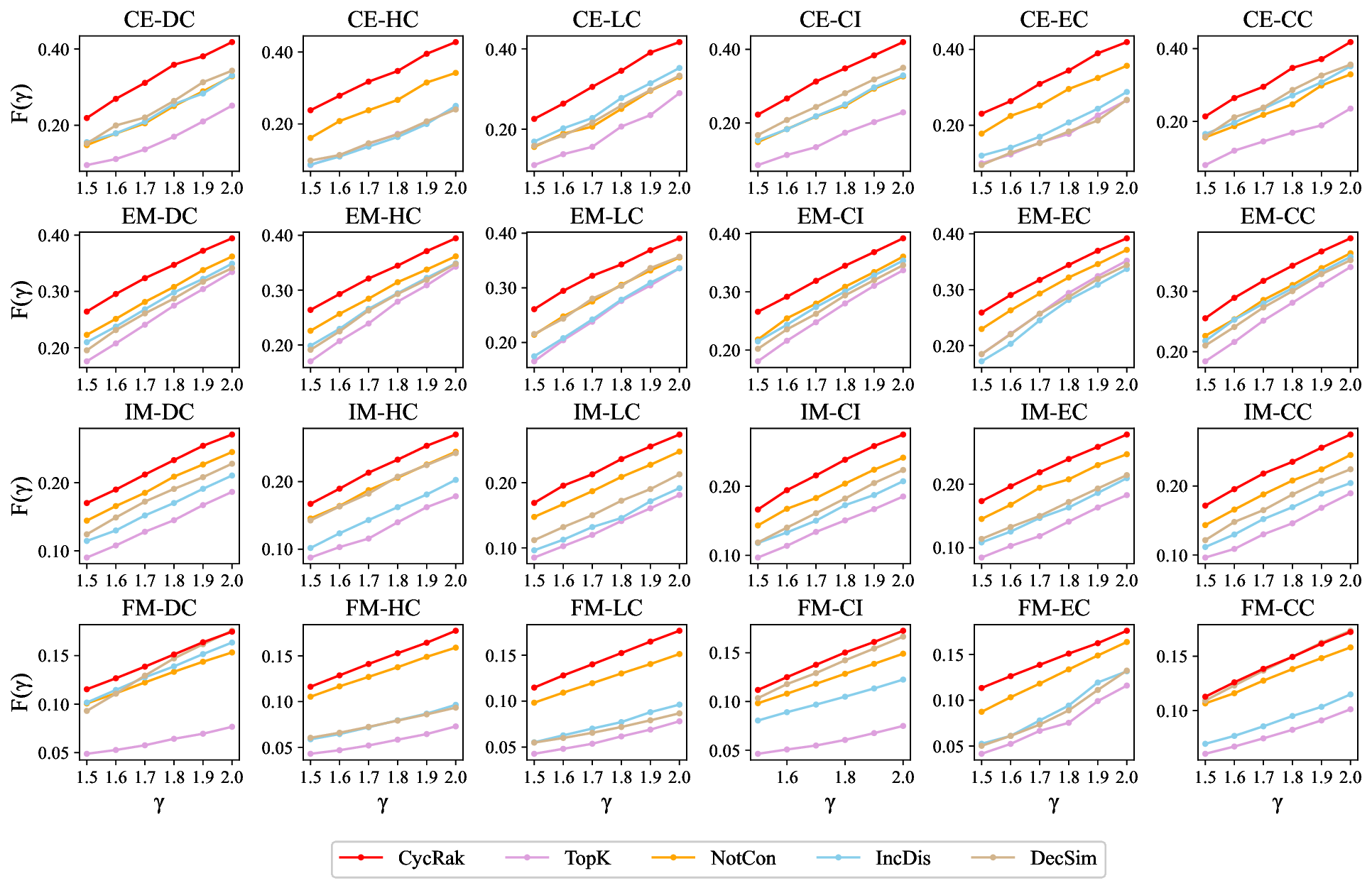}
\caption{Comparison of the influence with different infection probabilities on four empirical networks. Each panel presents results for a specific centrality, with the network and centrality name indicated in its title. Each curve depicts the collective influence $F(\gamma)$ of the top-2\% influencers selected by a particular optimization framework as a function of the infection probability $\gamma$. Here the x-axis is represented as the multiplier $\alpha \in [1.5,2.0]$ and $\gamma=\alpha \beta_c$. Each ($\gamma$, $F(\gamma)$) represents the average of 300 independent realizations, with $\mu=1$ held constant throughout.}
\label{fig4}
\end{figure}

\begin{figure}[!ht]
\centering
\includegraphics[width=0.7\textwidth, keepaspectratio]{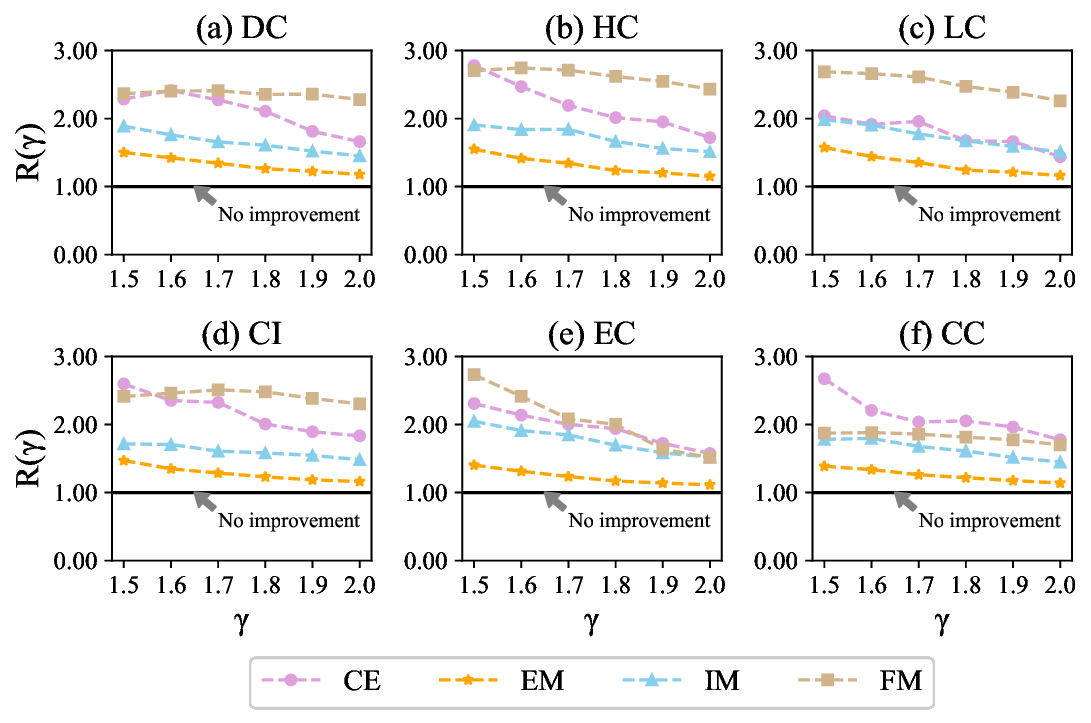}
\caption{Performance improvement of CycRak over TopK at different infection probabilities for four empirical networks. Here, $\gamma$ represents the infection probability and $\gamma=\alpha \beta_c$, where $\alpha \in [1.5,2.0]$. $R(\rho)$ indicates the factor by which the influence of the top-2\% influencers selected by CycRak exceeds that of TopK. The black solid line denotes the influence of TopK, serving as the baseline, with each dashed line corresponding to the results of CycRak in an empirical network, based on the centrality represented by the panel.}
\label{fig5}
\end{figure}

Figure~\ref{fig4} illustrates the influence of various frameworks under different infection probabilities $\gamma$ across four empirical networks. For brevity, the x-axis is depicted as the multiplier $\alpha \in [1.5,2.0]$, where $\gamma=\alpha \beta_c$ (holds true for Figures~\ref{fig5} and~\ref{fig9} as well). It can be observed that with increasing $\gamma$, the influence of all schemes steadily rises. However, regardless of the network or centrality metric, CycRak consistently outperforms other frameworks, maintaining this advantage across the entire range of $\gamma$. In contrast, TopK’s performance is consistently inferior. These findings align with those in Figure~\ref{fig2}, including the suboptimal performance of NotCon.

Figure~\ref{fig5} further quantifies the extent to which CycRak enhances the performance of TopK as depicted in Figure~\ref{fig4}. It can be observed that CycRak improves the performance across all centrality metrics, with the magnitude of improvement varying depending on the network, ranging from relatively minor enhancements in the EM network to predominantly 1.5 to 3 times improvements over TopK in the other networks. Particularly, the gains by CycRak are more pronounced at lower $\gamma$, diminishing as $\gamma$ increases, implying CycRak’s advantage in dispersibility. Comparing Figures~\ref{fig3} and ~\ref{fig5}, it is evident that CycRak's optimization prowess is more independent of centrality metrics and more dependent on network characteristics. Moreover, considering the significant differences in modularity $Q$~\cite{newman2004finding} among these four empirical networks, as shown in Table~\ref{tab2}, the comprehensive analysis of Figures~\ref{fig2} to ~\ref{fig5} reveals the robustness of CycRak’s performance with respect to the level of community structure.

\begin{figure}[htbp]
\centering
\includegraphics[width=\textwidth, keepaspectratio]{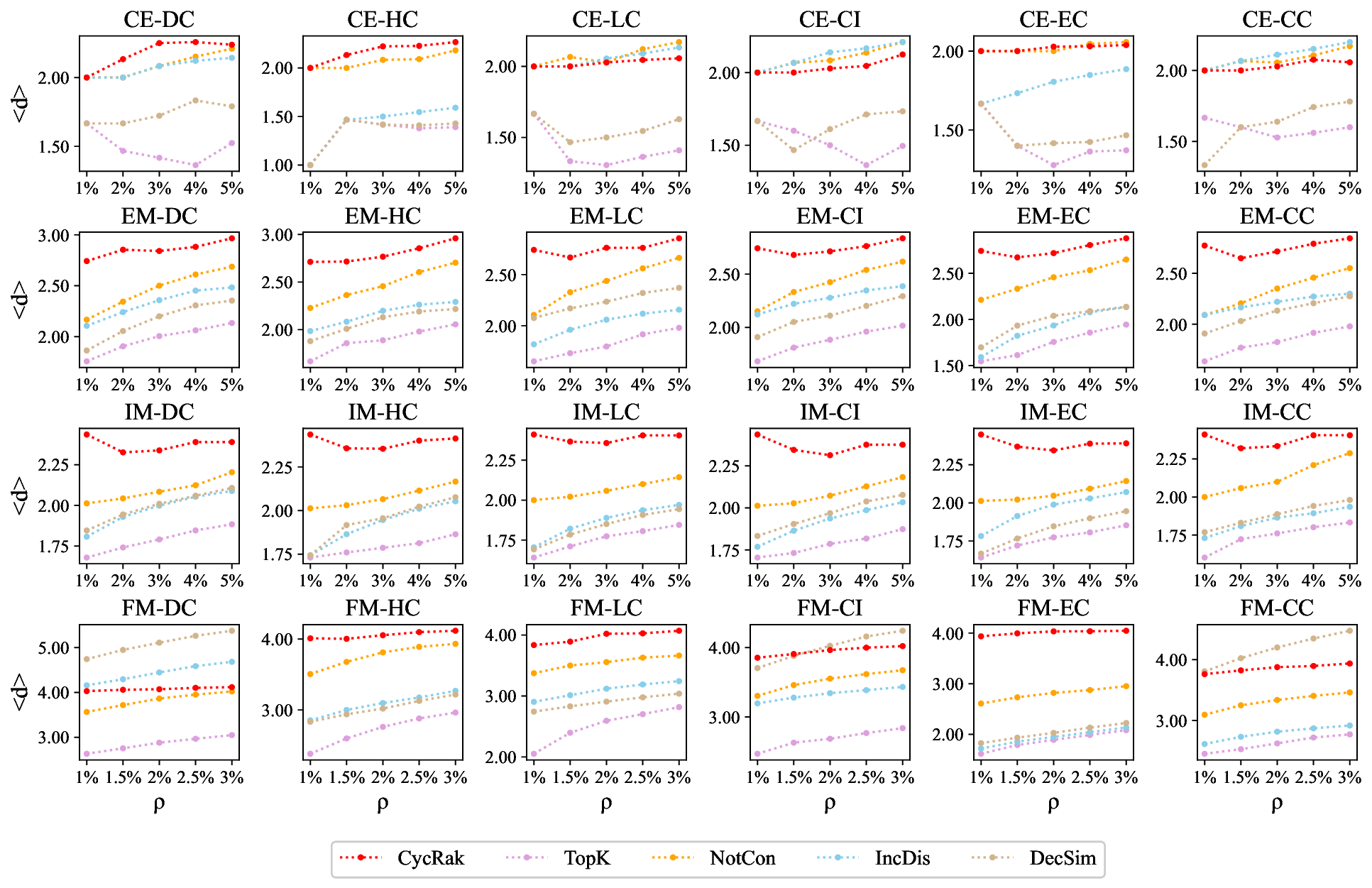}
\caption{Comparison of the average distance among influencers across four empirical networks. Here, $(\rho,\langle d \rangle)$ denotes that the average distance $\langle d \rangle$ among the top-$\rho$ influencers selected by the corresponding optimization framework. These results are captured under SIR model parameters $\gamma=1.25\beta_c$ and $\mu=1$.}
\label{fig6}
\end{figure}

Figure~\ref{fig6} elucidates one key reason for the superiority of CycRak, namely the greater distance between influencers. It can be observed that in the vast majority of cases, the average distance $\langle d \rangle$ between influencers selected by CycRak is significantly larger than that of other frameworks, notably the TopK method, by a factor of 1.25 to 2. This indicates that the influencers chosen by CycRak are more widely distributed throughout the entire network. Consequently, CycRak effectively avoids the overlap of the spheres of influence of influencers, thereby enhancing their collective effects. Furthermore, another advantage of CycRak compared to other frameworks is that $\langle d \rangle$ is less influenced by $\rho$, primarily due to CycRak prioritizing the ranking of influential cycles rather than centrality when selecting influencers. Additionally, the $\langle d \rangle$ of NotCon often ranks as the second largest, consistent with its suboptimal performance observed in previous results.

\begin{figure}[htbp]
\centering
\includegraphics[width=\textwidth, keepaspectratio]{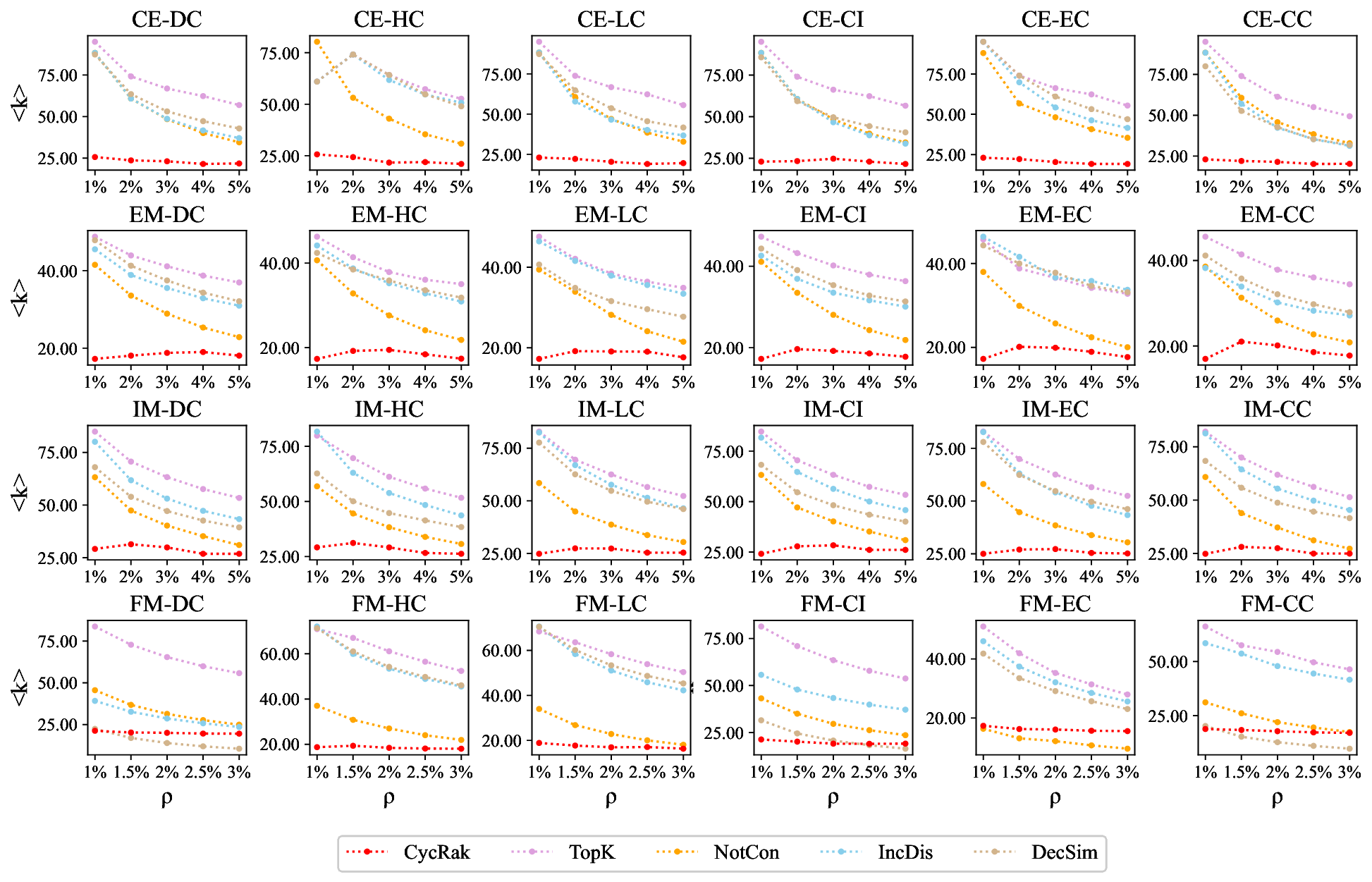}
\caption{Comparison of the average degree among influencers across four empirical networks. Here, $(\rho, \langle k \rangle)$ denotes that the average degree $\langle k \rangle$ among the top-$\rho$ influencers selected by the corresponding optimization framework. These results are captured under SIR model parameters $\gamma=1.25\beta_c$ and $\mu=1$.}
\label{fig7}
\end{figure}

Meanwhile, the average degree $\langle k \rangle$ of influencers selected by CycRak is the smallest, as shown in Figure~\ref{fig7}. Present-day research generally holds that influencers should exhibit both good dispersibility and as high a hub property as possible, reflected in a larger $\langle k \rangle$, as pursued by the four benchmark frameworks here. However, CycRak presents a counterexample, showcasing the weakest hub property, especially when $\rho$ is smaller, with $\langle k \rangle$ less than one-third of others. This is because, rather than selecting the centers of local clusters, which typically have higher degrees, as influencers, CycRak opts for nodes that hinge these local clusters together, which tend to have larger basic cycle numbers but limited degrees. This concept is achieved by Eq.~\ref{eq3}. Furthermore, CycRak’s $\langle k \rangle$ is barely influenced by $\rho$, while other frameworks significantly decrease with increasing $\rho$, primarily due to CycRak’s prioritization of cycle importance. Figure~\ref{fig7} illustrates that CycRak has opened up a new path toward maximizing influence, suggesting that more and better solutions are within reach. Additionally, smaller $\langle k \rangle$ often require smaller initial costs, thus Figure~\ref{fig7} also highlights CycRak’s excellent cost-effectiveness.

\subsection{Synthetic Networks}

\begin{figure}[htbp]
\centering
\includegraphics[width=\textwidth, keepaspectratio]{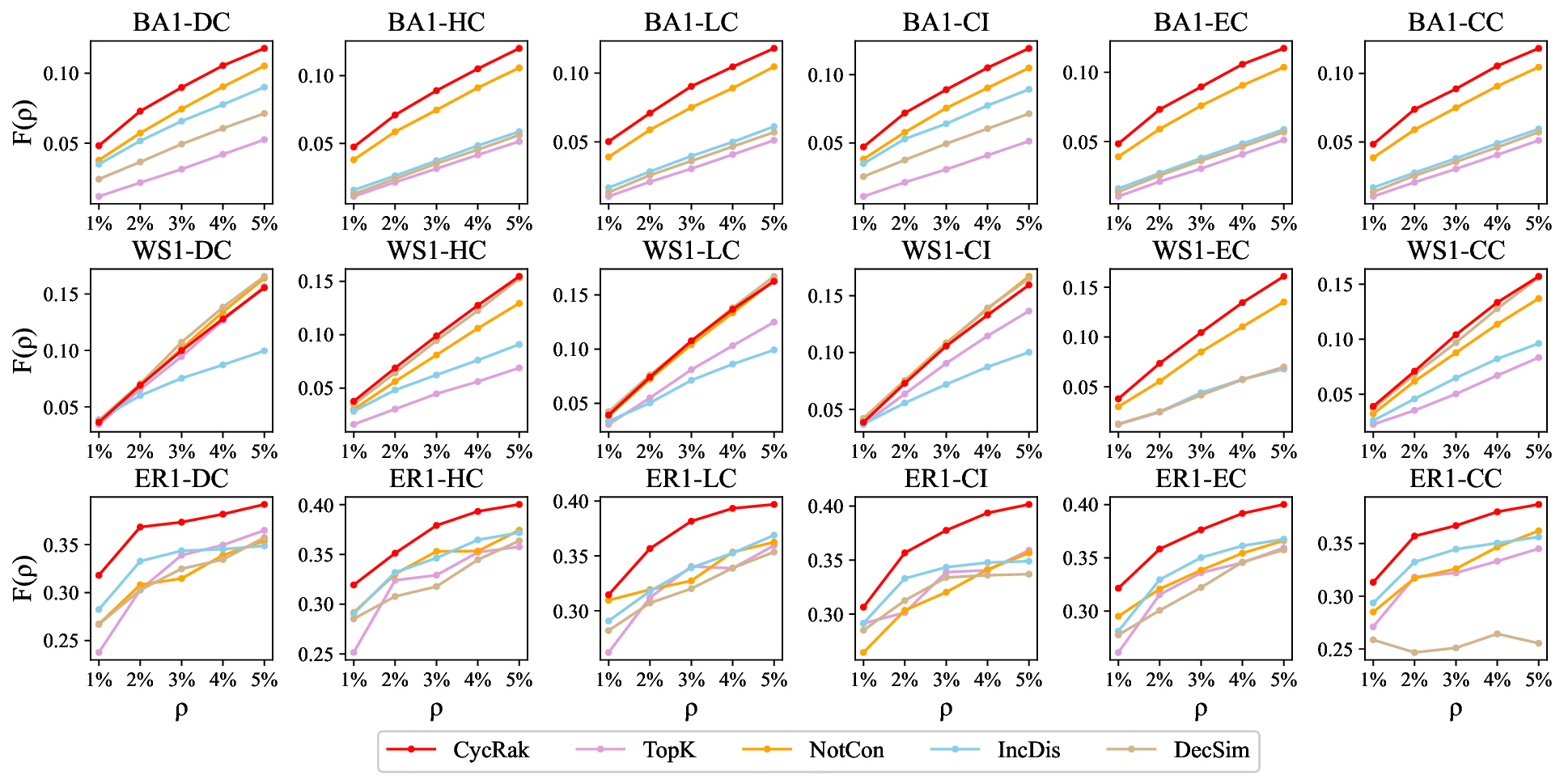}
\caption{Comparison of the influence across various influencer percentages for three synthetic networks. In networks BA1, WS1, and ER1, the SIR model’s $\gamma$ values are 0.05, 0.02, and 0.04, respectively, with $\mu$ fixed at 1. Each ($\rho$, $F(\rho)$) represents the average of 300 independent realizations.}
\label{fig8}
\end{figure}

\begin{figure}[htbp]
\centering
\includegraphics[width=\textwidth, keepaspectratio]{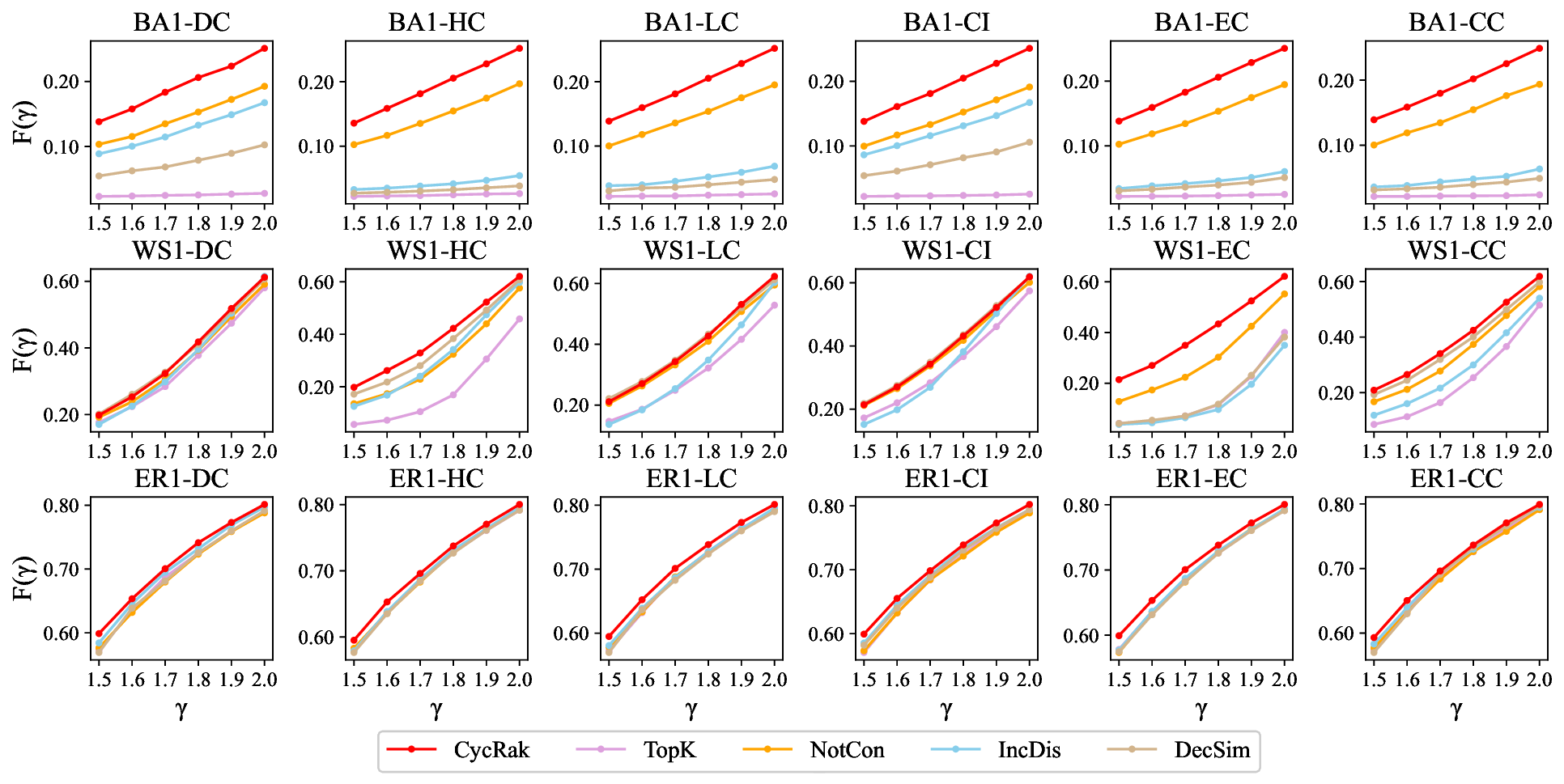}
\caption{Comparison of the influence with different infection probabilities for three synthetic networks. Here each curve depicts the influence $F(\gamma)$ of the top-2\% influencers selected by a particular optimization framework as a function of the infection probability $\gamma$, where $\gamma=\alpha \beta_c$ and $\alpha \in [1.5,2.0]$. Each ($\gamma$, $F(\gamma)$) represents the average of 300 independent realizations, with $\mu=1$.}
\label{fig9}
\end{figure}

\begin{figure}[htbp]
\centering
\includegraphics[width=\textwidth, keepaspectratio]{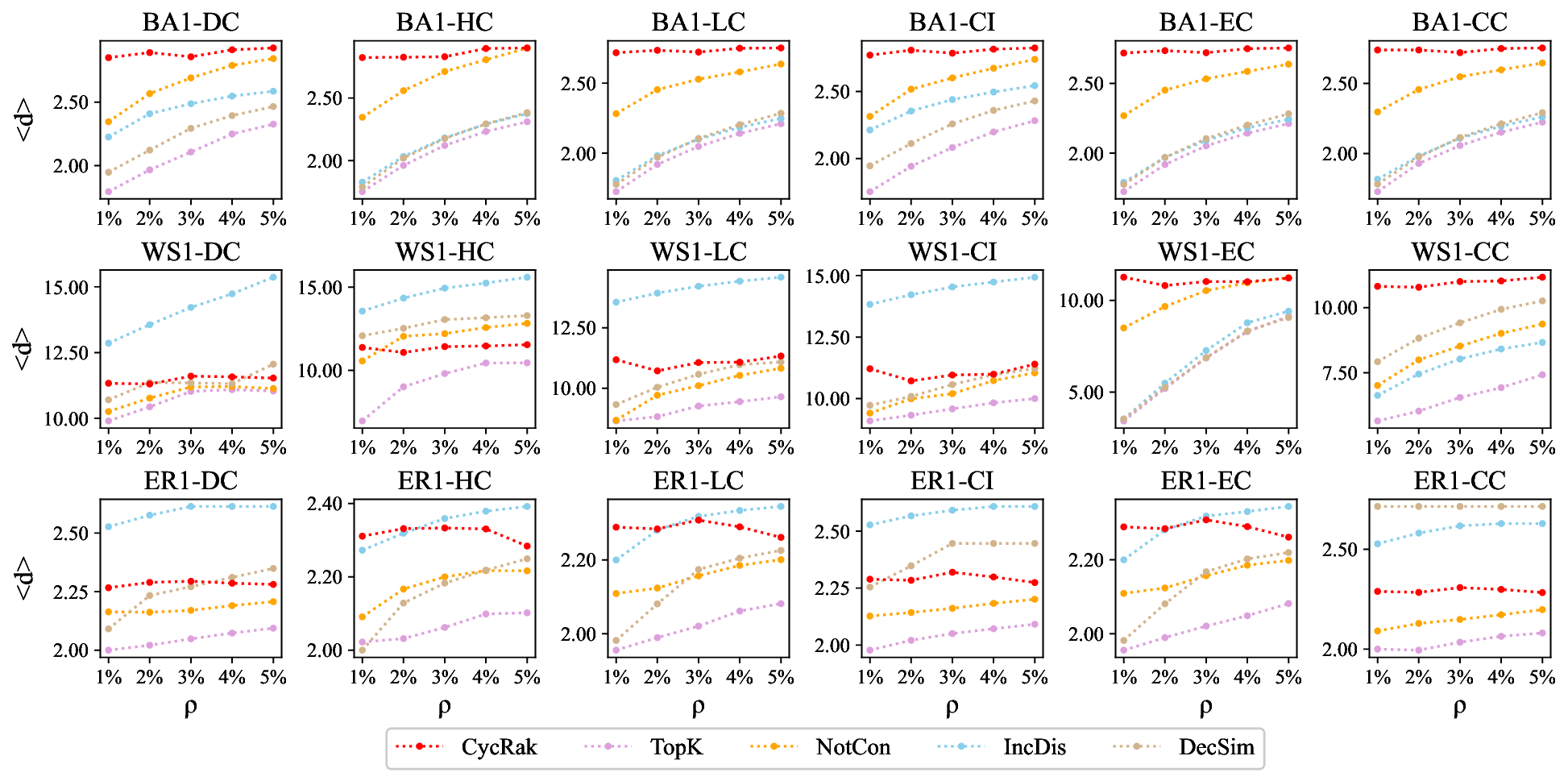}
\caption{Comparison of the average distance among influencers for three synthetic networks. Here, $(\rho,\langle d \rangle)$ denotes that the average distance $\langle d \rangle$ among the top-$\rho$ influencers selected by the corresponding optimization framework. In networks BA1, WS1, and ER1, the SIR model’s $\gamma$ values are 0.05, 0.02, and 0.04, respectively, with $\mu$ fixed at 1. Each $(\rho,\langle d \rangle)$ represents the average of 300 independent realizations.}
\label{fig10}
\end{figure}

\begin{figure}[htbp]
\centering
\includegraphics[width=\textwidth, keepaspectratio]{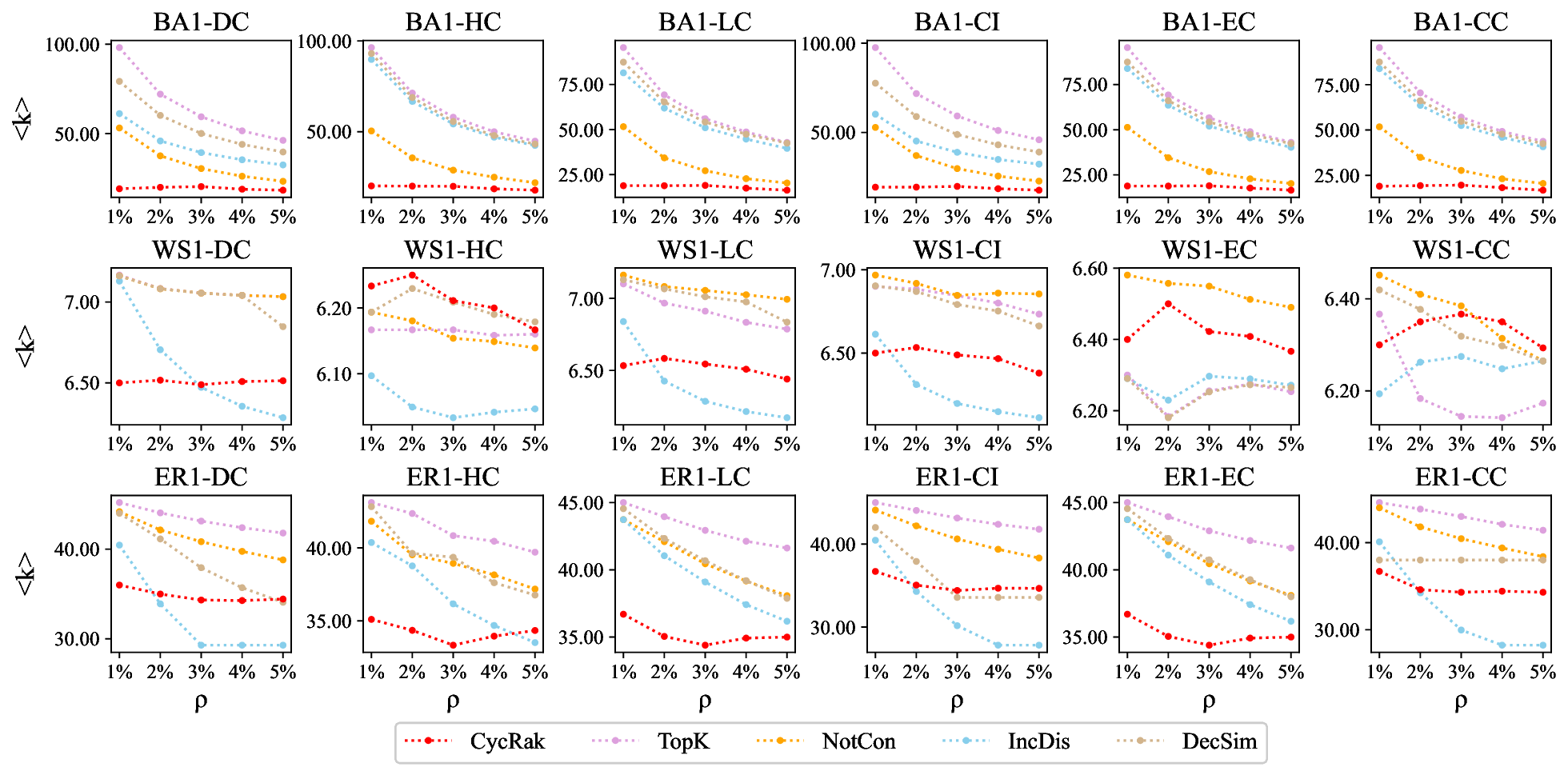}
\caption{Comparison of the average degree among influencers for three synthetic networks. Here, $(\rho, \langle k \rangle)$ denotes that the average degree $\langle k \rangle$ among the top-$\rho$ influencers selected by the corresponding optimization framework. In networks BA1, WS1, and ER1, the SIR model’s $\gamma$ values are 0.05, 0.02, and 0.04, respectively, with $\mu$ fixed at 1. Each $(\rho, \langle k \rangle)$ represents the average of 300 independent realizations.}
\label{fig11}
\end{figure}

Figures~\ref{fig8} to~\ref{fig11} illustrate the results of propagation experiments on three types of synthetic networks, the topological statistics of which are provided in Tabl~\ref{tab2}. Consistent with the findings on empirical networks, the influence of influencers selected by CycRak surpasses that of other benchmark frameworks, regardless of the influencers’ percentage, as depicted in Figure~\ref{fig8}, or the infection probability, as shown in Figure~\ref{fig9}. A novel observation in synthetic networks is that, in some cases, the optimized results obtained by employing NotCon, IncDis, and DecSim for TopK are inferior to the original TopK, with the exception of CycRak. This phenomenon is observed in the WS1 and ER1 networks depicted in Figure~\ref{fig8}. On the other hand, CycRak demonstrates a stronger ability to enhance influence in BA model networks, with influencers having greater distances between them and smaller average degrees. However, the degree of enhancement for WS and ER model networks is relatively small, with fluctuations in both dispersibility and low hub characteristic. This variation is attributed to the weaker emphasis on three features relied upon by CycRak, as characterized by Eqs~\ref{eq1} to ~\ref{eq3}, in these two types of networks.

Additional results can be found in Figures~\ref{fig:s2} to ~\ref{fig:s5} in \nameref{sec:SI}, which showcase the outcomes of six synthetic networks under different parameter configurations. These findings further confirm the consistent performance advantage of CycRak and suggest its potential across additional empirical or synthetic networks.

\section{Conclusions}

To optimize the performance of influence maximization strategies, we propose a new optimization framework called CycRak based on cycle ranking. CycRak utilizes key basic cycles to filter high-centrality nodes and capture influencers. Here, the importance of these basic cycles is characterized from three perspectives: their own significance, as well as the nodes and edges they comprise, from the aspects of community, effective paths, and articulation for the local network centers. This integration reflects the crucial role cycles play in networks from macro to micro levels. Experimental results demonstrate that the influencers identified by CycRak maximize propagation, outperforming all benchmark frameworks, with a dissemination range 1.5 to 3 times greater than other benchmarks, while exhibiting hub characteristics lower than one-third of the baseline. Additionally, they possess higher dispersibility, showcasing CycRak’s robust combinatorial effects and excellent cost advantages. The dominant advantage of CycRak remains stable regardless of the number of influencers, propagation probabilities, network sizes, or network types.

CycRak also has a unique advantage in that it can provide multiple alternative solutions with similar diffusion capabilities, attributed to the multiplicity of basic cycles, which endows CycRak with flexibility and adaptability. In addition to optimizing schemes based on centrality, CycRak theoretically can also optimize solutions obtained from other types of influence maximization algorithms. Furthermore, CycRak can also be directly used for the problem of identifying important cycles.

It is important to note that there is a significant difference between CycRak and other benchmark frameworks. While the benchmark methods primarily rely on centrality ranking, followed by individual screening based on specific rules, CycRak prioritizes cycle ranking, with centralities playing a secondary role. This fundamental difference is also the reason why the dispersibility and hub characteristics of CycRak are less affected by the number of influencers.

The advantages of CycRak demonstrate the significance of cycle structures in network dynamics and the immense potential in optimization problems. Unlike the meta-structures of networks such as nodes or edges, cycles, as multi-element objects, come in various sizes and can encapsulate more interactional information and patterns, exerting a more pronounced impact on the structure and dynamics of networks at the mesoscale level. Further exploration of the profound implications of cycle structures for networks represents an important task for future network research.

CycRak still has vast room for further expansion. On one hand, in terms of cycle types, CycRak only considers the basic cycles of the network. Other types of cycles, more complex and diverse, contain richer information and topology, potentially offering new insights. On the other hand, CycRak only deals with undirected and unweighted networks. Therefore, its extension and performance analysis on other types of networks, such as directed, weighted, or higher-order networks, are necessary and anticipated.

\vspace{\baselineskip}
\noindent \textbf{Code and data availability}. All codes and data supporting our findings are available from the GitHub repository: \url{https://github.com/xxx}.

\vspace{\baselineskip}
\noindent \textbf{Acknowledgments}. 
This research was funded by the National Natural Science Foundation of China (Grant No. T2293771), the support from the STI 2030--Major Projects (2022ZD0211400) and the New Cornerstone Science Foundation through the XPLORER PRIZE. The authors would like to thank Wenxin Zheng for her support in visualization.

\vspace{\baselineskip}
\noindent \textbf{Author contributions}. 
Wenfeng Shi, Tianlong Fan, Shuqi Xu and Linyuan Lü conceived the research and designed the experiments. Wenfeng Shi constructed the initial framework of the paper, carried out the numerical experiments, and drafted the manuscript. Rongmei Yang collected the literature and data, and supplemented the experiments. Tianlong Fan refined the design and experiments, and comprehensively revised the manuscript. Tianlong Fan, Shuqi Xu and Linyuan Lü supervised the research and performed the final manuscript editing. All authors reviewed and confirmed the methods and conclusions.

\vspace{\baselineskip}
\noindent \textbf{Competing interests}. 
The authors declare no competing interests.

\clearpage

\vspace{1.5cm}

\Acknowledgements{}

\bibliographystyle{unsrt}

\bibliography{references}

\newpage
\section*{Supplementary Information}

\addcontentsline{toc}{section}{Supplementary Information}\label{sec:SI}

\setcounter{figure}{0}
\setcounter{table}{0}
\setcounter{algorithm}{0}

\renewcommand{\thefigure}{\thesection.\arabic{figure}}
\renewcommand{\thefigure}{S\arabic{figure}} 

\renewcommand{\thetable}{\thesection.\arabic{table}}
\renewcommand{\thetable}{S\arabic{table}}

\renewcommand{\thealgorithm}{\thesection.\arabic{algorithm}}
\renewcommand{\thealgorithm}{S\arabic{algorithm}}

\begin{center}
\textbf{\large A universal optimization framework based on cycle ranking for influence maximization in complex networks}

\textbf{W Shi, T Fan, S Xu, R Yang \& L L\"u}
\end{center}

\begin{figure}[htbp]
\centering
\includegraphics[width=0.65\textwidth, keepaspectratio]{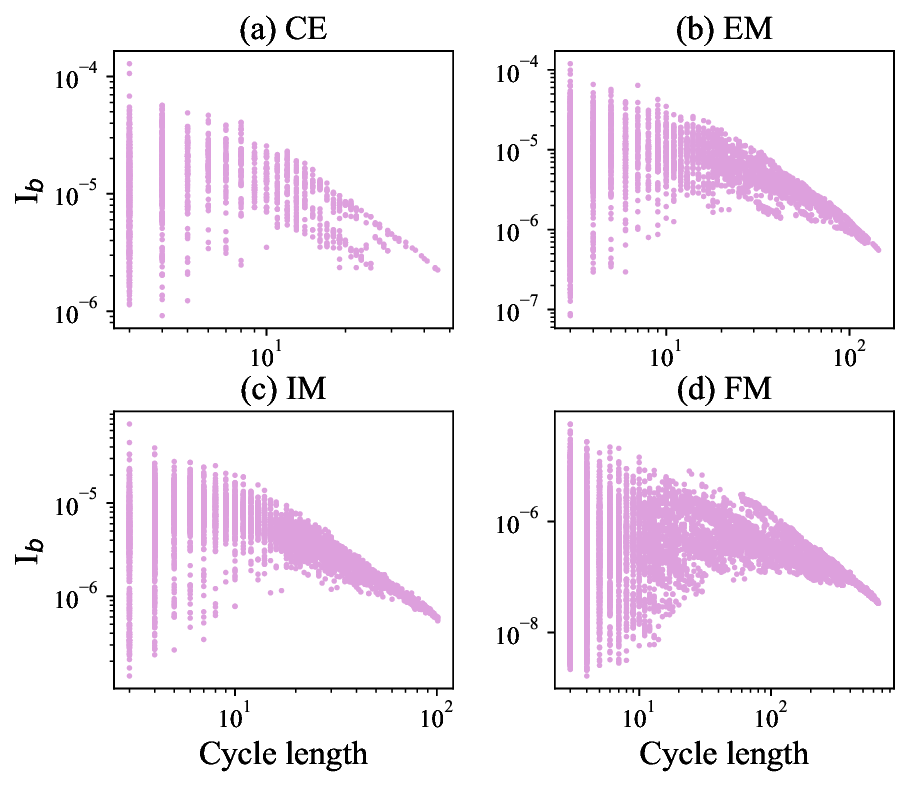}
\caption{Relationship between the importance $I_b$ of basic cycles and their cycle lengths (the edge number in the cycle) in four empirical networks.}
\label{fig:s1}
\end{figure}

\begin{table}[htbp]
\footnotesize
\caption{The importance $I_b$ and its components of the most important (Most Imp.) and least important (Least Imp.) basic cycles in four empirical networks.}
\tabcolsep 8pt
\centering
\begin{tabular}{ccccccccc}
\toprule
& \multicolumn{2}{c}{C. elegans (CE)} & \multicolumn{2}{c}{Email (EM)} & \multicolumn{2}{c}{Ia-fb-messages (IM)} & \multicolumn{2}{c}{Asian-last.fm (FM)} \\
\midrule
    & Most Imp. & Least Imp. & Most Imp. & Least Imp. & Most Imp. & Least Imp. & Most Imp. & Least Imp. \\
          % & (93,141,145) & (154,160,189,230) & (270,835,858) & (113,178,668) & (313,397,1100) & (831,1071,1074) & (1929,6252,7170) & (17,1737,2434,3165) \\
\midrule
    $I_{com}$ & 0.12765 & 0.02343 & 0.225 & 0.02777 & 0.21052 & 0.08333 & 0.14414 & 0.00532 \\
    $I_{pth}$ & 0.63663 & 0.0411 & 0.58324 & 0.00657 & 0.50272 & 0.00283 & 0.5177 & 0.00535 \\
    $I_{lc}$ & 0.00158 & 0.00095 & 0.00091 & 0.00045 & 0.00066 & 0.00059 & 0.00076 & 5.80E-05 \\
    $I_b$  & 0.00013 & 9.16E-07 & 0.00012 & 8.36E-08 & 7.06E-05 & 1.39E-07 & 5.72E-05 & 1.65E-09 \\
    \bottomrule
    \end{tabular}%
  \label{tab:s1}%
\end{table}%

% \begin{center}
% \begin{minipage}[c]{0.9\textwidth}
\begin{algorithm}[H]
\renewcommand{\algorithmicrequire}{\textbf{Input:}}
\renewcommand{\algorithmicensure}{\textbf{Output:}}
\footnotesize
\caption{\textbf{Pseudocode for the cycle-ranking based framework for identifying influencers (\emph{CycRak}).}}

\label{alg1}
\begin{algorithmic}[1]
    \REQUIRE Graph $G(V, E)$, node centrality dictionary $C_{en}$;
    \ENSURE  A set of multiple influential nodes, $S$;
    \STATE Initialization: $P \leftarrow \varnothing$, $S \leftarrow \varnothing$
    \STATE Find a set of basic cycles $B$ in the network $G$
    \FOR{$b$ in $B$}
        \STATE Calculate $I_{com}$ value of $b$
        \STATE Calculate $I_{pth}$ value of $b$
        \STATE Calculate $I_{lc}$ value of $b$
        \STATE Calculate $I_b$ value of $b$
    \ENDFOR
    \STATE Rank all basic cycles by $I_b$ value to obtain the cycle ranking list $R_c$
    \WHILE{$|S| < k$}
        \STATE Select the top-ranked cycle $b$ in $R_c$, $b' = b - S$
        \STATE Find centrality values for all nodes in $b'$ from $C_{en}$
        \STATE Pick the node $i$ with the highest centrality in $b'$ and add it to $P$
        \IF{$i$ is not directly connected to any node in $S$}
            \STATE Add $i$ to $S$
            \STATE Remove $b$ from $R_c$
        \ELSE
            \STATE Remove $b$ from $R_c$
        \ENDIF
    \ENDWHILE
\end{algorithmic}
\end{algorithm}
% \end{minipage}
% \end{center}

\begin{figure}[htbp]
\centering
\includegraphics[width=\textwidth, keepaspectratio]{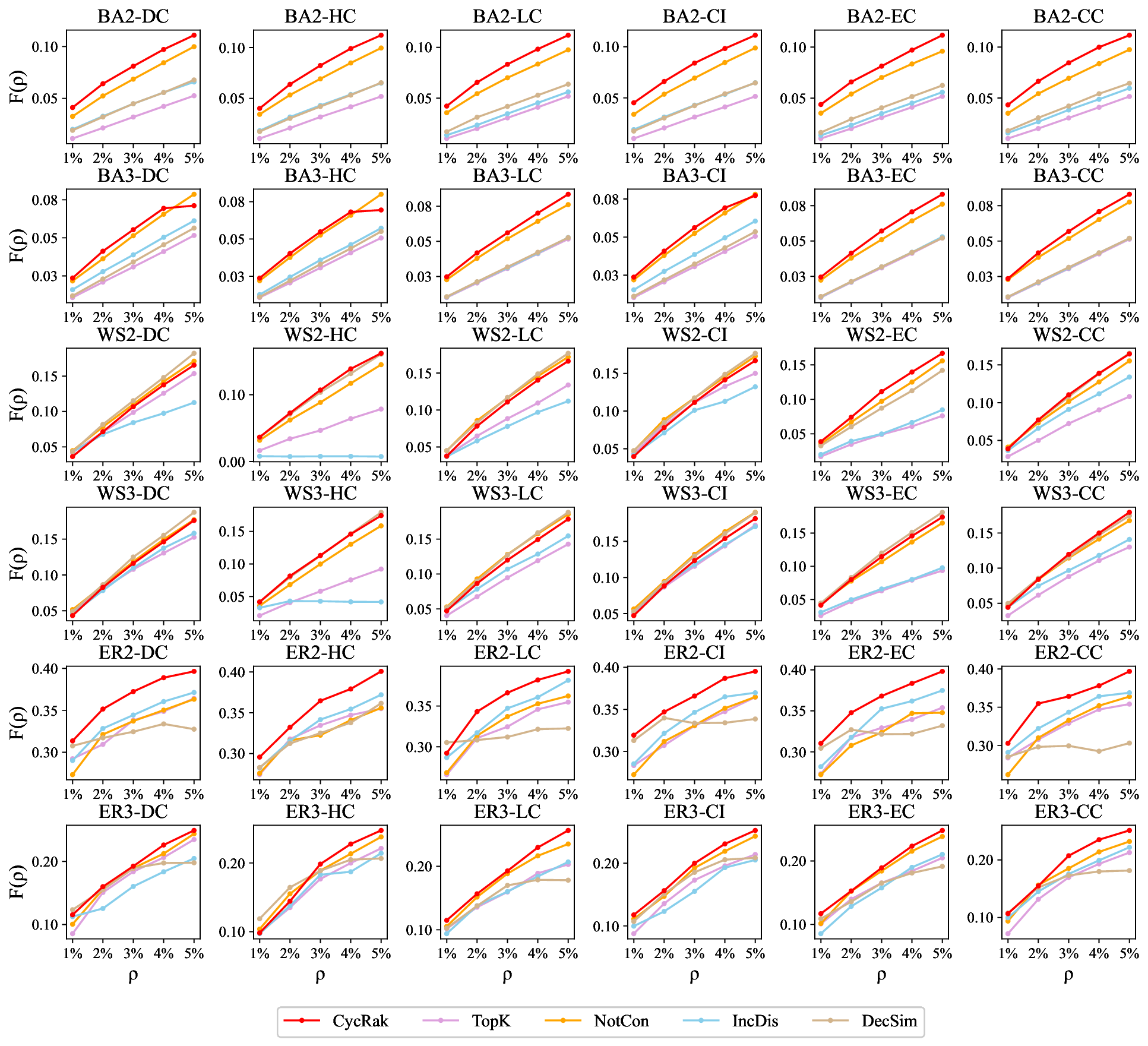}
\caption{Comparison of the influence across various influencer percentages for the other six synthetic networks. In networks BA2, BA3, WS2, WS3, ER2, and ER3, the SIR model’s $\gamma$ values are 0.06, 0.07, 0.02, 0.02, 0.06, and 0.1, respectively, with $\mu$ fixed at 1. Each ($\rho$, $F(\rho)$) represents the average of 300 independent realizations.}
\label{fig:s2}
\end{figure}

\begin{figure}[htbp]
\centering
\includegraphics[width=\textwidth, keepaspectratio]{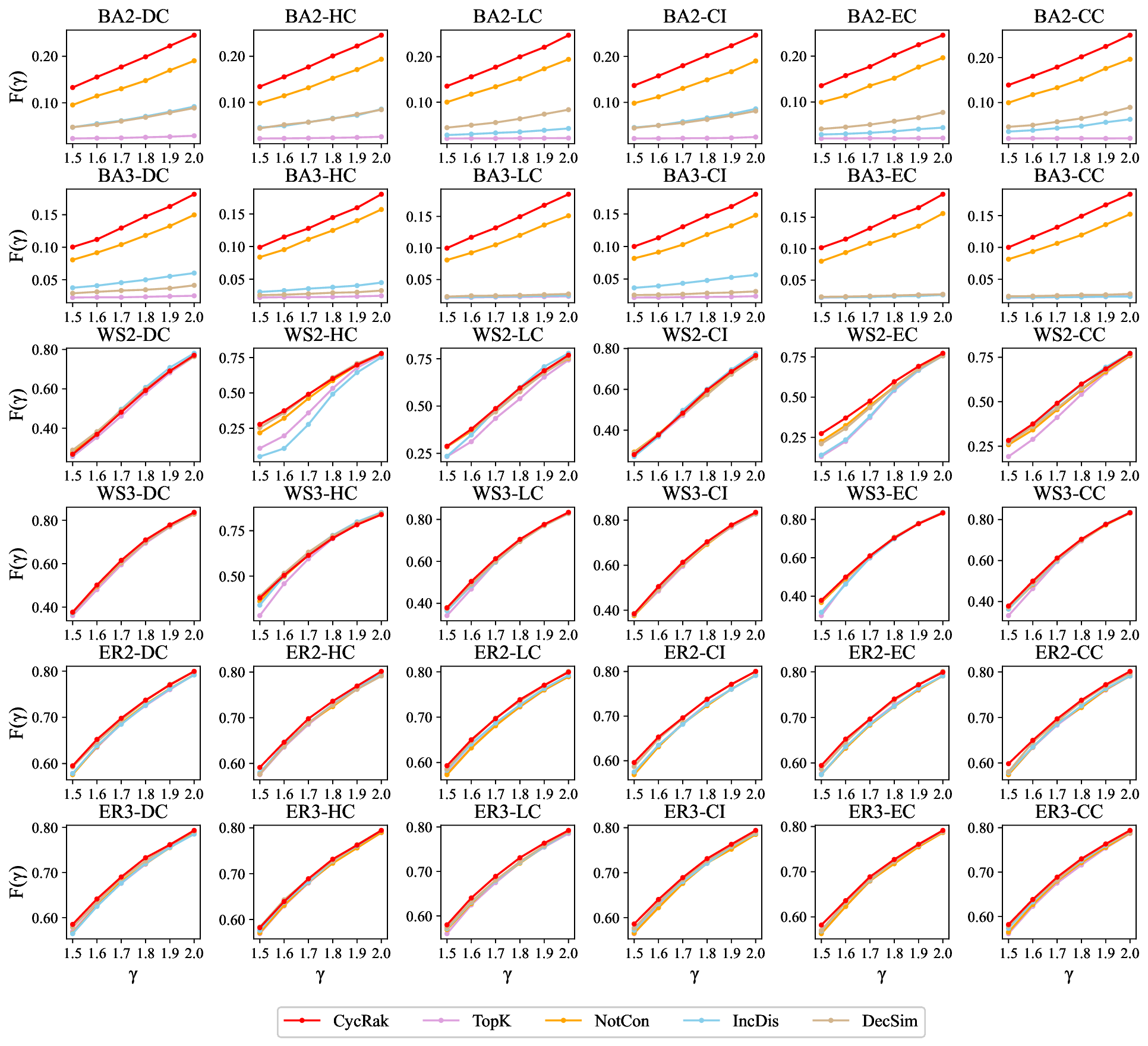}
\caption{Comparison of the influence with different infection probabilities for the other six synthetic networks. Here each curve depicts the influence $F(\gamma)$ of the top-2\% influencers selected by a particular optimization framework as a function of the infection probability $\gamma$, where $\gamma=\alpha \beta_c$ and $\alpha \in [1.5,2.0]$. Each ($\gamma$, $F(\gamma)$) represents the average of 300 independent realizations, with $\mu=1$.}
\label{fig:s3}
\end{figure}

\begin{figure}[htbp]
\centering
\includegraphics[width=\textwidth, keepaspectratio]{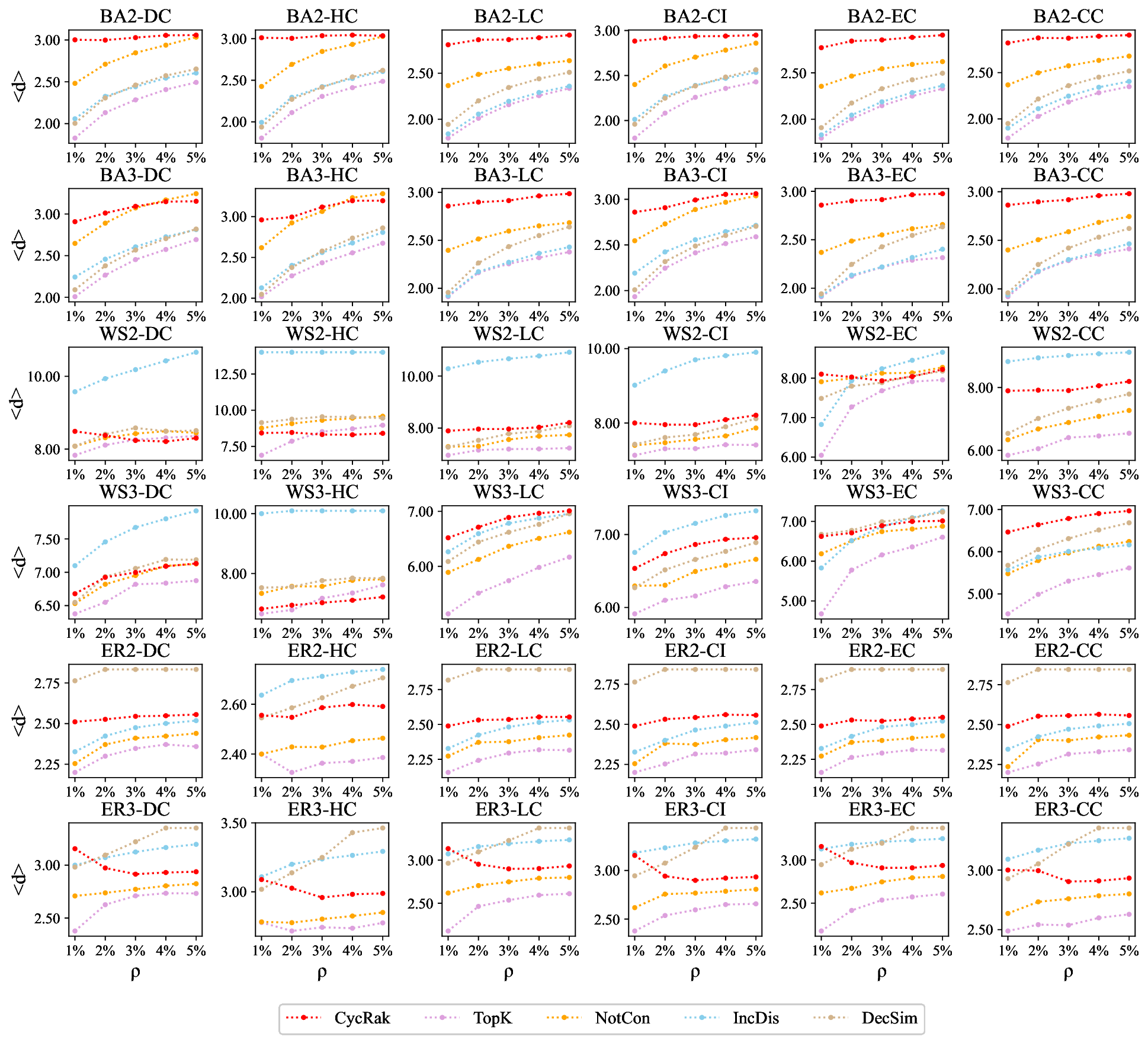}
\caption{Comparison of the average distance among influencers for the other six synthetic networks. Here, $(\rho,\langle d \rangle)$ denotes that the average distance $\langle d \rangle$ among the top-$\rho$ influencers selected by the corresponding optimization framework. In networks BA2, BA3, WS2, WS3, ER2, and ER3, the SIR model’s $\gamma$ values are 0.06, 0.07, 0.02, 0.02, 0.06, and 0.1, respectively, with $\mu$ fixed at 1. Each $(\rho,\langle d \rangle)$ represents the average of 300 independent realizations.}
\label{fig:s4}
\end{figure}

\begin{figure}[htbp]
\centering
\includegraphics[width=\textwidth, keepaspectratio]{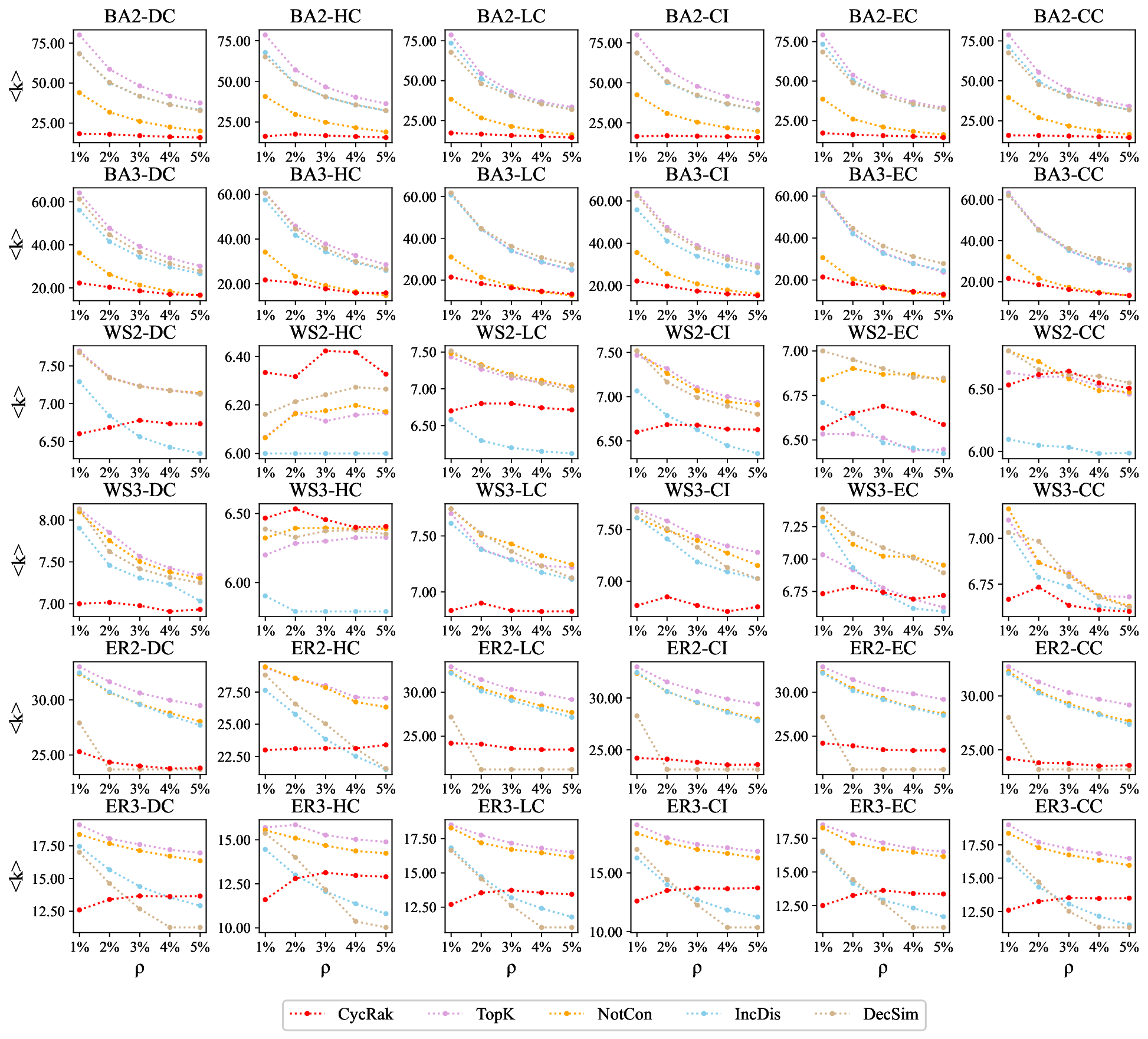}
\caption{Comparison of the average degree among influencers for the other six synthetic networks. Here, $(\rho, \langle k \rangle)$ denotes that the average degree $\langle k \rangle$ among the top-$\rho$ influencers selected by the corresponding optimization framework. In networks BA2, BA3, WS2, WS3, ER2, and ER3, the SIR model’s $\gamma$ values 0.06, 0.07, 0.02, 0.02, 0.06, and 0.1, respectively, with $\mu$ fixed at 1. Each $(\rho, \langle k \rangle)$ represents the average of 300 independent realizations.}
\label{fig:s5}
\end{figure}

\end{document}